\documentclass[showpacs,aps,graphicx,twocolumn]{revtex4}
\usepackage{amsmath}
\usepackage{mathrsfs}
\usepackage{amsfonts}
\usepackage{amssymb}
\usepackage{graphicx}
\usepackage{subfigure}
\usepackage{eufrak}
\usepackage{multirow}
\usepackage{longtable}
\usepackage{float}
\usepackage{bm}
\usepackage{xcolor}
\usepackage[colorlinks,linkcolor=blue,anchorcolor=blue,citecolor=blue,urlcolor=blue]{hyperref}
\usepackage{soul,color,xcolor}

\begin{document}

\title{Practically Enhanced Hyperentanglement Concentration for Polarization-spatial Hyperentangled Bell States with Linear Optics and Common Single-photon Detectors}

\author{Gui-Long Jiang,\textsuperscript{1} Wen-Qiang Liu,\textsuperscript{2} and Hai-Rui Wei\textsuperscript{1,}}
\email[]{hrwei@ustb.edu.cn}
\address{\textsuperscript{\rm1} School of Mathematics and Physics, University of Science and Technology Beijing, Beijing 100083, China\\
\textsuperscript{\rm2} Center for Quantum Technology Research and Key Laboratory of Advanced Optoelectronic Quantum
Architecture and Measurements (MOE), School of Physics, Beijing Institute of Technology, Beijing 100081, China}

\date{Received 22 September 2022; revised 10 December 2022; accepted 13 February 2023; published 14 March 2023}

\begin{abstract}

Hyperentanglement, defined as the simultaneous entanglement in several independent degrees of freedom (DOFs) of a quantum system, is a fascinating resource in quantum information processing with its outstanding merits.
Here we propose heralded hyperentanglement concentration protocols (hyper-ECPs) to concentrate an unknown partially less polarization-spatial hyperentangled Bell state with available linear optics and common single-photon detectors.
By introducing time-delay DOFs, the schemes are highly efficient in that the success of the scheme can be accurately heralded by the detection signatures, and postselection techniques or photon-number-resolving detectors, necessary for previous experiments, are not required.
Additionally, our linear optical architectures allow certain states, where concentration fails, to be recyclable, and a trick makes the success probabilities of our schemes higher than those of previous linear optical hyper-ECPs.


\end{abstract}


\maketitle

\section{Introduction}\label{sec1}

Entanglement, as an extremely characteristic quantum mechanical phenomenon, plays a significant part in quantum information processing and has been widely applied to quantum communication and quantum computation \cite{1,2}.
The entangled photons, which can carry quantum information through various qubitlike degrees of freedom (DOFs), have been recognized as the optimum candidates in long-distance quantum communication tasks, such as quantum key distribution \cite{3,4,5}, quantum teleportation \cite{6,7}, quantum dense coding \cite{8,9}, quantum secret sharing \cite{10,11,12}, and quantum secure direct communication \cite{13,14,15}.
Encoding qubits in the polarization DOF of photons has been particularly appealing, for their arbitrary single-qubit manipulation can be easily achieved with two quarter-wave plates and one half-wave plate in an extremely fast and accurate manner \cite{single-qubit}.
Recently, hyperentanglement, defined as the simultaneous entanglement in multiple independent DOFs of a quantum system, has captured much attention as it can considerably increase the capacity of quantum communication and speed up quantum computation \cite{17,18,momentum1,momentum2,OAM1,OAM2,OAM3,HPQC1,HPQC2,HPQC3,HBSA1}.
Diverse kinds of hyperentanglement have been experimentally demonstrated in detail, e.g., polarization-spatial \cite{17}, polarization-spatial-time \cite{18}, polarization-momentum \cite{momentum1,momentum2}, and polarization-orbital-angular-momentum \cite{OAM1,OAM2} DOFs.

The hyperentangled states can break the channel capacity limit of superdense coding \cite{OAM3} and provide substantial applications in hyperparallel optical quantum computing \cite{HPQC1,HPQC2,HPQC3} and hyperparallel communication, such as hyperentanglement swapping \cite{HBSA1,HBSA2}, hyperentangled Bell states analysis \cite{17,HBSA1,HBSA2,HBSA3,HBSA4,HBSA-wei}, and hyperentanglement purification \cite{HEPP1,HEPP2,HEPP3}.
Moreover, hyperentangled states can be used to accomplish some specific tasks that are impossible to achieve in single-DOF systems, such as complete and deterministic Bell states analysis with linear optics \cite{momentum2,BSA1,BSA2,BSA3}.
However, the entangled photons will inevitably be affected by environment noise during transmission and storage processes in long-distance quantum communication, resulting in a great probability of degradation of the fidelity and security of protocols.
One effective method to prevent the degradation of entanglement in photon systems is entanglement concentration protocol (ECP), which can distill maximally entangled states from less-entangled pure states. Another approach is entanglement purification protocol (EPP), which is applied to extract maximally entangled states with high fidelity from a number of mixed entangled states \cite{EPP1,EPP2,EPP3,EPP4,EPP5}.
In 1996, Bennett \emph{et al.} \cite{ECP1} proposed an ECP by using the Schmidt projection method, and various interesting ECPs and improvements were later proposed and experimentally demonstrated \cite{ECP2,ECP2-experiment1,ECP3,ECP4,ECP5,ECP6,ECP7}.

The current ECPs are mostly focused on single-DOF systems.
In 2013, Ren \emph{et al.} \cite{hyper-ECP1} proposed an interesting polarization-spatial hyper-ECP for known hyperentangled Bell states with linear optics by using the parameter-splitting structure.
In 2015, Li and Ghose \cite{hyper-ECP2} presented two hyper-ECPs for hyperentangled states in the polarization and time-bin DOFs with known and unknown parameters by using linear elements with similar methods.
Later, Ren and Wang \emph{et al.} \cite{hyper-ECP3} designed polarization-spatial-time-bin hyper-ECPs for  hyperentangled Bell states and  hyperentangled Greenberger-Horne-Zeilinger (GHZ) states \cite{hyper-ECP4} with linear optical elements.
In 2019, Li and Shen \cite{hyper-ECP02} reported asymmetrical hyper-ECPs for unknown and known photon systems entangled in polarization and orbital angular momentum DOFs.
For less-entangled states with known parameters, parameter splitting is the current optimal method for entanglement concentration with linear optics \cite{hyper-ECP1}.
While for less-entangled states with unknown parameters,  polarization beam splitters (PBSs) are usually employed to complete parity-check measurement on the polarization photon pair \cite{hyper-ECP2,hyper-ECP3,hyper-ECP4,hyper-ECP02}, and postselections or sophisticated photon-number-resolving detectors  are necessary to accomplish the scheme exactly with linear optics.
In addition, some deterministic hyper-ECPs were proposed assisted by nonlinear platforms, such as cross-Kerr mediates \cite{hyper-ECP5}, quantum dots \cite{hyper-ECP6}, or diamond nitrogen-vacancy centers \cite{hyper-ECP7,hyper-ECP8}, and these methods are challenged by impracticality in experiments with the current technique.

In this paper, we present two practical hyper-ECPs for partially hyperentangled states with unknown parameters.
A maximally polarization-spatial-based hyperentangled Bell (GHZ) state is exactly extracted from a partially less hyperentangled Bell (GHZ) state by using linear optics and common single-photon detectors.
Using the method of introducing temporal DOF in Ref. \cite{BSA3}, we orchestrate unbalanced interferences that allow postselection principles or sophisticated photon-number-resolving detectors to be avoided and the success of the schemes to be completely heralded by the detection signatures.
As the heralded concentration equipment is composed only of linear optical elements and has less apparatus than the previous schemes, our schemes are quite applicable and practical with current technology.
Besides, the success probability of our hyper-ECPs can be higher than that of the existing hyper-ECPs with linear optics.


%
%


\section{Hyper-ECP for photon systems in unknown hyperentangled pure states}\label{sec2}

In this section, we construct heralded hyper-ECPs for partially less hyperentangled Bell states and GHZ states in both polarization and spatial DOFs with unknown parameters by using linear optics.
The basic principles of our hyper-ECPs for the hyperentangled Bell and GHZ states are shown in Fig. \ref{Fig.1} and Fig. \ref{Fig.3}, respectively.



\subsection{Hyper-ECP for partially hyperentangled Bell states with unknown parameters}\label{sec2.1}

Assume that two initial two-photon states $|\phi\rangle_{AB}$ and $|\phi\rangle_{A^{\prime}B^{\prime}}$ partially entangled in polarization and spatial DOFs simultaneously are generated from $S_{1}$ and $S_{2}$, respectively.
Here $|\phi\rangle_{AB}$ and $|\phi\rangle_{A^{\prime}B^{\prime}}$ are given by the following forms:
\begin{eqnarray}\label{eq1}
\begin{split}
&|\phi\rangle_{AB}=\frac{1}{2}(|HH\rangle+|VV\rangle) \otimes (|a_{1}b_{1}\rangle+|a_{2}b_{2}\rangle),\\
&|\phi\rangle_{A^{\prime}B^{\prime}}=\frac{1}{2}(|HH\rangle+|VV\rangle) \otimes (|a_{1^{\prime}}b_{1^{\prime}}\rangle+|a_{2^{\prime}}b_{2^{\prime}}\rangle).
\end{split}
\end{eqnarray}
where $H$ ($V$) denotes the horizontal (vertical) polarized photon. The $|a_{1}\rangle$ and $|a_{2}\rangle$ ($|a_{1^{\prime}}\rangle$ and $|a_{2^{\prime}}\rangle$) are two spatial modes of photon $A$ ($A^{\prime}$) and the $|b_{1}\rangle$ and $|b_{2}\rangle$ ($|b_{1^{\prime}}\rangle$ and $|b_{2^{\prime}}\rangle$) are those of photon $B$ ($B^{\prime}$).
Then, an initial four-photon state of the four-photon system consisting of photons $A$, $B$, $A^{\prime}$, and $B^{\prime}$ is given by
\begin{eqnarray}\label{eq2}
\begin{split}
|\Phi_{0}\rangle
=&|\phi\rangle_{AB} \otimes |\phi\rangle_{A^{\prime}B^{\prime}}\\
=&\frac{1}{4}(|HH\rangle+|VV\rangle) \otimes (|HH\rangle+|VV\rangle) \\
&\otimes (|a_{1}b_{1}\rangle+|a_{2}b_{2}\rangle) \otimes (|a_{1^{\prime}}b_{1^{\prime}}\rangle+|a_{2^{\prime}}b_{2^{\prime}}\rangle).
\end{split}
\end{eqnarray}

\begin{figure*} [htbp]
  \centering
  \includegraphics[width=18 cm]{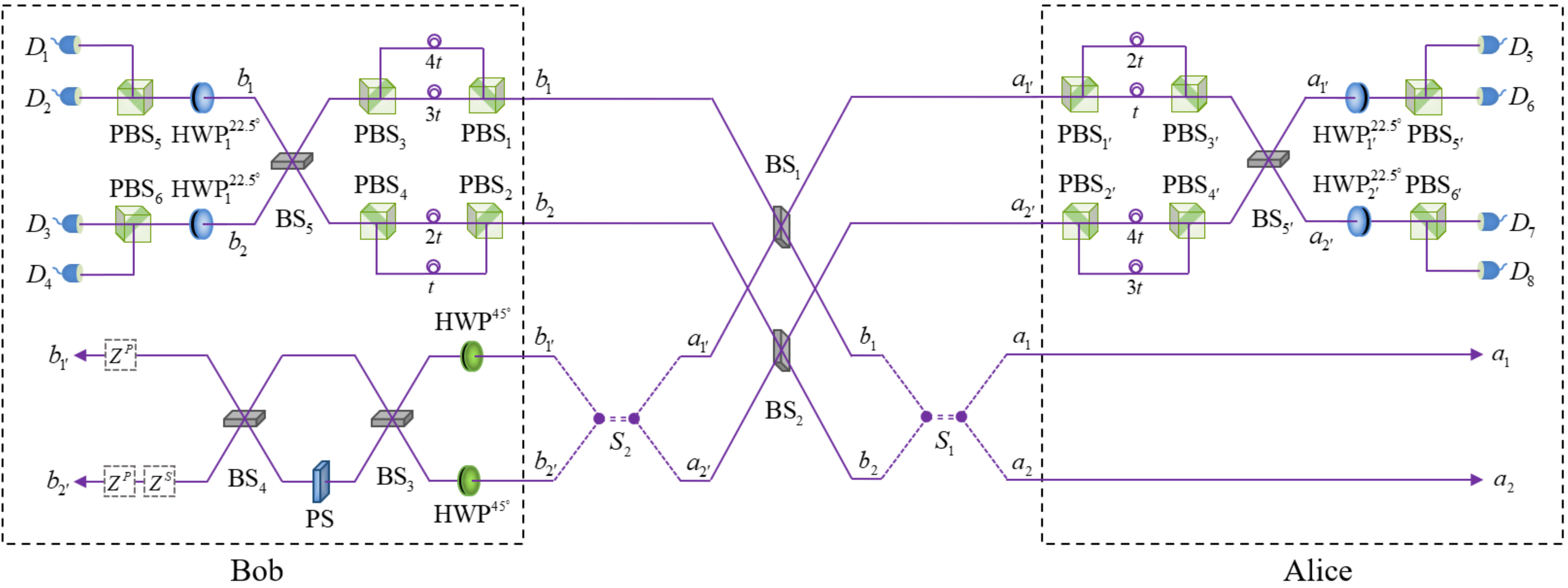}
  \caption{(Color online) Schematic diagram of the hyper-ECP for a polarization-spatial-based hyperentangled Bell state with unknown parameters.
  $S_{1}$ and $S_{2}$ are partial hyperentanglement sources for $|\phi\rangle_{AB}$ and $|\phi\rangle_{A^{\prime} B^{\prime}}$, respectively.
  $\textrm{PBS}_{i,i^{\prime}}(i=1,2,\cdots,6)$ represents a polarization beam splitter, which transmits the horizontal polarization state $|H\rangle$ and reflects the vertical polarization state $|V\rangle$.
  $\textrm{HWP}^{45^{\circ}}$ represents a half-wave plate oriented at $45^{\circ}$, which is used to exchange the vertical and horizontal polarization, i.e., $|H\rangle \rightarrow |V\rangle$, $|V\rangle \rightarrow |H\rangle$.
  $\textrm{HWP}^{22.5^{\circ}}_{i}(i=1,2)$ is used to perform a Hadamard operation on polarization, i.e., $|H\rangle \rightarrow \frac{1}{\sqrt{2}}(|H\rangle+|V\rangle)$, $|V\rangle \rightarrow \frac{1}{\sqrt{2}}(|H\rangle-|V\rangle)$.
  $\textrm{BS}_{i,i^{\prime}}(i=1,2,\cdots,5)$ denotes the 50:50 beam splitter.
  $D_{i}$ $(i=1,2,\cdots,8)$ represents the common single-photon detector.
  $Z^{S}$ and $Z^{P}$ are the classical feed-forward operations, which can be achieved by setting a PS and a $\textrm{HWP}^{0^{\circ}}$, respectively.}\label{Fig.1}
\end{figure*}

As shown in Fig. \ref{Fig.1}, photon $B$ in the spatial mode $|b_{1}\rangle$ ($|b_{2}\rangle$) and photon $A^{\prime}$ in the spatial mode $|a_{1^{\prime}}\rangle$ ($|a_{2^{\prime}}\rangle$) first pass through a 50:50 beam splitter BS$_{1}$ (BS$_{2}$), where the balanced BS$_{1}$ or BS$_{2}$ matrix is given by
\begin{eqnarray}\label{eq3}
\text{BS}=\frac{1}{\sqrt{2}}\left(
  \begin{array}{cc}
    1 & 1  \\
    1 & -1 \\
  \end{array}
\right)
\end{eqnarray}
in the spatial mode $\{|b_i\rangle, |a_{i^{\prime}}\rangle\}$ ($i=1,2$) basis.
After passing through BS$_{1}$ and BS$_{2}$, $|\Phi_{0}\rangle$ is transformed into
\begin{eqnarray}\label{eq4}
\begin{split}
|\Phi_{1}\rangle
=&\frac{1}{8}(|HH\rangle+|VV\rangle) \otimes (|HH\rangle+|VV\rangle) \\
&\otimes [|a_{1}\rangle(|b_{1}\rangle+|a_{1^{\prime}}\rangle)+|a_{2}\rangle(|b_{2}\rangle+|a_{2^{\prime}}\rangle)]\\
&\otimes [(|b_{1}\rangle-|a_{1^{\prime}}\rangle)|b_{1^{\prime}}\rangle+(|b_{2}\rangle-|a_{2^{\prime}}\rangle)|b_{2^{\prime}}\rangle].
\end{split}
\end{eqnarray}
Then, photons $A$ and $A^{\prime}$ ($B$ and $B^{\prime}$) are sent to Alice (Bob), and owing to the influence of noise, $|\Phi_{1}\rangle$ may decay to a partially less entangled state
\begin{eqnarray}\label{eq5}
\begin{split}
|\Phi_{2}\rangle
=&\frac{1}{2}(\alpha|HH\rangle+\beta|VV\rangle) \otimes (\alpha|HH\rangle+\beta|VV\rangle) \\
&\otimes [\gamma|a_{1}\rangle(|b_{1}\rangle+|a_{1^{\prime}}\rangle)+\delta|a_{2}\rangle(|b_{2}\rangle+|a_{2^{\prime}}\rangle)]\\
&\otimes [\gamma(|b_{1}\rangle-|a_{1^{\prime}}\rangle)|b_{1^{\prime}}\rangle+\delta(|b_{2}\rangle-|a_{2^{\prime}}\rangle)|b_{2^{\prime}}\rangle].
\end{split}
\end{eqnarray}
Here coefficients satisfy the normalization conditions $|\alpha|^{2}+|\beta|^{2}=1$ and $|\gamma|^{2}+|\delta|^{2}=1$, and they are unknown to the parties of communication.

The purpose of the two distant parties is to distill the maximally hyperentangled Bell state
\begin{eqnarray}\label{eq6}
\begin{split}
|\phi^{++}\rangle_{AB}=\frac{1}{2}(|HH\rangle+|VV\rangle) \otimes (|a_{1}b_{1}\rangle+|a_{2}b_{2}\rangle)
\end{split}
\end{eqnarray}
from $|\Phi_{2}\rangle$.
For clarity of exposition of the working principles of our ECP, base on the Hong-Ou-Mandel effect, we rewrite $|\Phi_{2}\rangle$ in the following normalized form:
\begin{eqnarray}\label{eq7}
\begin{split}
|\Phi_{2}\rangle
=&\frac{1}{\sqrt{2}}(\alpha^{2}|HHHH\rangle+\beta^{2}|VVVV\rangle)\\
&\otimes  [\gamma^{2}|a_{1}b_{1^{\prime}}\rangle(|b_{1}b_{1}\rangle-|a_{1^{\prime}}a_{1^{\prime}}\rangle)\\
&+\delta^{2}|a_{2}b_{2^{\prime}}\rangle(|b_{2}b_{2}\rangle-|a_{2^{\prime}}a_{2^{\prime}}\rangle)]\\
&+ \frac{\alpha\beta}{2}(|HVHV\rangle+|VHVH\rangle)\\
&\otimes  [\gamma^{2}|a_{1}b_{1^{\prime}}\rangle(|b_{1}b_{1}\rangle-|a_{1^{\prime}}a_{1^{\prime}}\rangle-|b_{1}a_{1^{\prime}}\rangle\\&
+|a_{1^{\prime}}b_{1}\rangle)+\delta^{2}|a_{2}b_{2^{\prime}}\rangle(|b_{2}b_{2}\rangle-|a_{2^{\prime}}a_{2^{\prime}}\rangle\\&
-|b_{2}a_{2^{\prime}}\rangle+|a_{2^{\prime}}b_{2}\rangle)]\\
&+ \frac{\gamma\delta}{2}(\alpha^{2}|HHHH\rangle+\beta^{2}|VVVV\rangle\\
&+\alpha\beta|HVHV\rangle+\beta\alpha|VHVH\rangle)\\
&\otimes  [|a_{1}b_{2^{\prime}}\rangle(|b_{1}b_{2}\rangle-|a_{1^{\prime}}a_{2^{\prime}}\rangle-|b_{1}a_{2^{\prime}}\rangle+|a_{1^{\prime}}b_{2}\rangle)\\
&+|a_{2}b_{1^{\prime}}\rangle(|b_{2}b_{1}\rangle-|a_{2^{\prime}}a_{1^{\prime}}\rangle-|b_{2}a_{1^{\prime}}\rangle+|a_{2^{\prime}}b_{1}\rangle)].
%
\end{split}
\end{eqnarray}

As shown in Fig. \ref{Fig.1}, Bob flips firstly the polarization and spatial states of photon $B^{\prime}$, which causes $|\Phi_{2}\rangle$ to change into
%
\begin{eqnarray}\label{eq8}
\begin{split}
|\Phi_{3}\rangle
=&\frac{1}{\sqrt{2}}(\alpha^{2}|HVHH\rangle+\beta^{2}|VHVV\rangle)\\
&\otimes  [\gamma^{2}|a_{1}b_{2^{\prime}}\rangle(|b_{1}b_{1}\rangle-|a_{1^{\prime}}a_{1^{\prime}}\rangle)\\
&+\delta^{2}|a_{2}b_{1^{\prime}}\rangle(|b_{2}b_{2}\rangle-|a_{2^{\prime}}a_{2^{\prime}}\rangle)]\\
&+ \frac{\alpha\beta}{2}(|HHHV\rangle+|VVVH\rangle)\\
&\otimes  [\gamma^{2}|a_{1}b_{2^{\prime}}\rangle(|b_{1}b_{1}\rangle-|a_{1^{\prime}}a_{1^{\prime}}\rangle-|b_{1}a_{1^{\prime}}\rangle\\&
+|a_{1^{\prime}}b_{1}\rangle)+\delta^{2}|a_{2}b_{1^{\prime}}\rangle(|b_{2}b_{2}\rangle-|a_{2^{\prime}}a_{2^{\prime}}\rangle\\&
-|b_{2}a_{2^{\prime}}\rangle+|a_{2^{\prime}}b_{2}\rangle)]\\
&+ \frac{\gamma\delta}{2}(\alpha^{2}|HVHH\rangle+\beta^{2}|VHVV\rangle\\
&+\alpha\beta|HHHV\rangle+\beta\alpha|VVVH\rangle)\\
&\otimes  [|a_{1}b_{1^{\prime}}\rangle(|b_{1}b_{2}\rangle-|a_{1^{\prime}}a_{2^{\prime}}\rangle-|b_{1}a_{2^{\prime}}\rangle+|a_{1^{\prime}}b_{2}\rangle)\\
&+|a_{2}b_{2^{\prime}}\rangle(|b_{2}b_{1}\rangle-|a_{2^{\prime}}a_{1^{\prime}}\rangle-|b_{2}a_{1^{\prime}}\rangle+|a_{2^{\prime}}b_{1}\rangle)],
%
\end{split}
\end{eqnarray}
where the polarization-based bit-flip operation is realized by two half-wave plates oriented at $45^{\circ}$ (HWP$^{45^{\circ}}$s), and the spatial-based bit-flip operation is realized by the block composed of BS$_3$, BS$_4$, and one phase shifter (PS).
%
%
The PS and the HWP$^{45^{\circ}}$ matrices are given by
\begin{eqnarray}             \label{eq9}
\text{PS}=\left(
  \begin{array}{cc}
    -1 & 0  \\
    0 & -1 \\
  \end{array}
\right),\;\;
\text{HWP}^{45^{\circ}}=\left(
  \begin{array}{cc}
    0 & 1  \\
    1 & 0 \\
  \end{array}
\right),
\end{eqnarray}
in the $\{|H\rangle, |V\rangle\}$ basis, respectively.


Secondly, as shown in Fig. \ref{Fig.1}, by using four unbalanced interferometers, each consisting of two PBSs, Alice introduces time delays
$t$  to $|H\rangle|a_{1^{\prime}}\rangle$,
$2t$ to $|V\rangle|a_{1^{\prime}}\rangle$,
$3t$ to $|V\rangle|a_{2^{\prime}}\rangle$, and
$4t$ to $|H\rangle|a_{2^{\prime}}\rangle$, respectively.
While Bob introduces time-delays
$t$  to  $|V\rangle|b_2\rangle$,
$2t$ to  $|H\rangle|b_2\rangle$,
$3t$ to  $|H\rangle|b_1\rangle$, and
$4t$ to  $|V\rangle|b_1\rangle$.
The time delays can be achieved by lengthening optical circuits, and we assume that phase $\omega t=  2 n \pi$, where $n$ is the nonzero integer and the time for photons to pass through the entire device is much less than the two-photon coherence time.
We can verify that after photons $A^{\prime}$ and $B$ are emitted from the four interferometers, $|\Phi_{3}\rangle$ becomes
\begin{eqnarray}\label{eq10}
|\Phi_{4}\rangle=|\Omega_{0}\rangle+|\Omega_{1}\rangle+|\Omega_{2}\rangle+|\Omega_{3}\rangle,
\end{eqnarray}
where
\begin{eqnarray}\label{eq11}
\begin{split}
|\Omega_{0}\rangle=&\frac{\alpha\beta}{2}[\gamma^{2}(|HH\rangle |a_{1}b_{2^{\prime}}\rangle+|VV\rangle |a_{1}b_{2^{\prime}}\rangle)\\&
\otimes (|HV\rangle |b_{1}b_{1}\rangle |3t, 4t\rangle-|HV\rangle |a_{1^{\prime}}a_{1^{\prime}}\rangle |t, 2t\rangle)\\&
+\delta^{2}(|HH\rangle |a_{2}b_{1^{\prime}}\rangle+|VV\rangle |a_{2}b_{1^{\prime}}\rangle)\\&
\otimes (|HV\rangle |b_{2}b_{2}\rangle |2t, t\rangle-|HV\rangle |a_{2^{\prime}}a_{2^{\prime}}\rangle |4t, 3t\rangle)\\&
+\gamma^{2}(|HH\rangle |a_{1}b_{2^{\prime}}\rangle-|VV\rangle |a_{1}b_{2^{\prime}}\rangle)\\&
\otimes (|HV\rangle |a_{1^{\prime}} b_{1}\rangle |t, 4t\rangle-|HV\rangle |b_{1} a_{1^{\prime}}\rangle |3t, 2t\rangle)\\&
+\delta^{2}(|HH\rangle |a_{2}b_{1^{\prime}}\rangle-|VV\rangle |a_{2}b_{1^{\prime}}\rangle)\\&
\otimes (|HV\rangle |a_{2^{\prime}} b_{2}\rangle |4t, t\rangle-|HV\rangle |b_{2} a_{2^{\prime}}\rangle |2t, 3t\rangle)],
\end{split}
\end{eqnarray}
\begin{eqnarray}\label{eq12}
\begin{split}
|\Omega_{1}\rangle=&\frac{\gamma\delta}{2}[\alpha^{2}(|HV\rangle |a_{1}b_{1^{\prime}}\rangle+|HV\rangle |a_{2}b_{2^{\prime}}\rangle)\\&
\otimes (|HH\rangle |b_{1}b_{2}\rangle |3t, 2t\rangle-|HH\rangle |a_{1^{\prime}}a_{2^{\prime}}\rangle |t, 4t\rangle)\\&
+\alpha^{2}(|HV\rangle |a_{1}b_{1^{\prime}}\rangle-|HV\rangle |a_{2}b_{2^{\prime}}\rangle)\\&
\otimes (|HH\rangle |a_{1^{\prime}}b_{2}\rangle |t, 2t\rangle-|HH\rangle |b_{1}a_{2^{\prime}}\rangle |3t, 4t\rangle)\\&
+\beta^{2}(|VH\rangle |a_{1}b_{1^{\prime}}\rangle+|VH\rangle |a_{2}b_{2^{\prime}}\rangle)\\&
\otimes (|VV\rangle |b_{1} b_{2}\rangle |4t, t\rangle-|VV\rangle |a_{1^{\prime}} a_{2^{\prime}}\rangle |2t, 3t\rangle)\\&
+\beta^{2}(|VH\rangle |a_{1}b_{1^{\prime}}\rangle-|VH\rangle |a_{2}b_{2^{\prime}}\rangle)\\&
\otimes (|VV\rangle |a_{1^{\prime}} b_{2}\rangle |2t, t\rangle-|VV\rangle |b_{1} a_{2^{\prime}}\rangle |4t, 3t\rangle)],
\end{split}
\end{eqnarray}
\begin{eqnarray}\label{eq13}
\begin{split}
|\Omega_{2}\rangle=&\frac{\alpha\beta\gamma\delta}{2}[(|HH\rangle |a_{1}b_{1^{\prime}}\rangle+|VV\rangle |a_{2}b_{2^{\prime}}\rangle)\\&
\otimes (|HV\rangle |b_{1}b_{2}\rangle |3t, t\rangle-|HV\rangle |a_{1^{\prime}}a_{2^{\prime}}\rangle |t, 3t\rangle)\\&
+(|HH\rangle |a_{2}b_{2^{\prime}}\rangle+|VV\rangle |a_{1}b_{1^{\prime}}\rangle)\\&
\otimes (|VH\rangle |b_{1}b_{2}\rangle |4t,2t\rangle-|VH\rangle |a_{1^{\prime}}a_{2^{\prime}}\rangle |2t, 4t\rangle)\\&
+(|HH\rangle |a_{1}b_{1^{\prime}}\rangle-|VV\rangle |a_{2}b_{2^{\prime}}\rangle)\\&
\otimes (|HV\rangle |a_{1^{\prime}} b_{2}\rangle |t, t\rangle-|HV\rangle |b_{1} a_{2^{\prime}}\rangle |3t, 3t\rangle)\\&
+(|HH\rangle |a_{2}b_{2^{\prime}}\rangle-|VV\rangle |a_{1}b_{1^{\prime}}\rangle)\\&
\otimes (|HV\rangle |a_{2^{\prime}} b_{1}\rangle |4t, 4t\rangle-|HV\rangle |b_{2} a_{1^{\prime}}\rangle |2t, 2t\rangle)].
\end{split}
\end{eqnarray}
\begin{eqnarray}\label{eq14}
\begin{split}
|\Omega_{3}\rangle
     =&\frac{1}{\sqrt{2}}[\alpha^{2}\gamma^{2}|HV\rangle |a_{1}b_{2^{\prime}}\rangle\\&
     \otimes (|HH\rangle |b_{1}b_{1}\rangle |3t,3t\rangle-|HH\rangle |a_{1^{\prime}}a_{1^{\prime}}\rangle |t,t\rangle)\\&
     +\beta^{2}\gamma^{2}|VH\rangle |a_{1}b_{2^{\prime}}\rangle \\&
     \otimes (|VV\rangle |b_{1}b_{1}\rangle |4t,4t\rangle-|VV\rangle |a_{1^{\prime}}a_{1^{\prime}}\rangle |2t,2t\rangle)\\&
     +\alpha^{2}\delta^{2}|HV\rangle |a_{2}b_{1^{\prime}}\rangle \\&
     \otimes (|HH\rangle |b_{2}b_{2}\rangle |2t,2t\rangle-|HH\rangle |a_{2^{\prime}}a_{2^{\prime}}\rangle |4t,4t\rangle)\\&
     +\beta^{2}\delta^{2}|VH\rangle |a_{2}b_{1^{\prime}}\rangle \\&
     \otimes (|VV\rangle |b_{2}b_{2}\rangle |t,t\rangle-|VV\rangle |a_{2^{\prime}}a_{2^{\prime}}\rangle |3t,3t\rangle)],
\end{split}
\end{eqnarray}
Based on Eqs. (\ref{eq11}-\ref{eq14}), we can see that $|\Omega_{0}\rangle$ and $|\Omega_{1}\rangle$ indicate the failure of the protocol, that is, the scheme is terminated; $|\Omega_{2}\rangle$ has the potential to yield the maximal hyperentangled Bell state since it has the same coefficient for each item, whereas $|\Omega_{3}\rangle$ is promising for recycling to improve the success probability as it can yield a partially hyperentangled Bell state \cite{hyper-ECP5,hyper-ECP6,hyper-ECP7}.
The rigorous argument is provided below.

%

Thirdly, Alice (Bob) performs spatial-based and polarization-based Hadamard operations on photon $A^{\prime}$ ($B$) by $\text{BS}_{5^{\prime}}$ and $\text{HWP}_{1^\prime,2^\prime}^{22.5^{\circ}}$ ($\text{BS}_5$ and $\text{HWP}_{1,2}^{22.5^{\circ}}$), respectively.
Here the $\text{HWP}^{22.5^{\circ}}$ matrix is given by
\begin{eqnarray}\label{eq15}
\text{HWP}^{22.5^{\circ}}=\frac{1}{\sqrt{2}}\left(
  \begin{array}{cc}
    1 & 1  \\
    1 & -1 \\
  \end{array}
\right),
\end{eqnarray}
in the $\{|H\rangle, |V\rangle\}$ basis.
Here, $\text{BS}_{5^{\prime}}$ and $\text{BS}_5$ complete the transformations
\begin{eqnarray}\label{eq16}
\begin{split}
&|a_{1^{\prime}}\rangle \rightarrow \frac{1}{\sqrt{2}}(|a_{1^{\prime}}\rangle+|a_{2^{\prime}}\rangle),\;\;
|b_{1}\rangle \rightarrow \frac{1}{\sqrt{2}}(|b_{1}\rangle+|b_{2}\rangle),\\&
|a_{2^{\prime}}\rangle \rightarrow \frac{1}{\sqrt{2}}(|a_{1^{\prime}}\rangle-|a_{2^{\prime}}\rangle),\;\;
|b_{2}\rangle \rightarrow \frac{1}{\sqrt{2}}(|b_{1}\rangle-|b_{2}\rangle).
\end{split}
\end{eqnarray}
For the desired state $|\Omega_{2}\rangle$ shown in Eq. (\ref{eq13}), after photon $A^{\prime}$ ($B$) passes through $\text{BS}_{5^{\prime}}$ ($\text{BS}_5$) and $\text{HWP}_{1^\prime,2^\prime}^{22.5^{\circ}}$ ($\text{HWP}_{1,2}^{22.5^{\circ}}$) in succession, if both photons have the same polarization, spatial mode, and time interval, they will interfere  constructively or destructively with each other, e.g., $|HH\rangle |a_{1^{\prime}}b_{1}\rangle |t, t\rangle - |HH\rangle |a_{1^{\prime}}b_{1}\rangle |3t, 3t\rangle \rightarrow 0$, $|HV\rangle |a_{1^{\prime}}b_{1}\rangle |t, t\rangle+|VH\rangle |b_{1}a_{1^{\prime}}\rangle |3t, 3t\rangle \rightarrow 2|HV\rangle |a_{1^{\prime}}b_{1}\rangle D(0,0)$, and $|HV\rangle |b_{1}b_{2}\rangle |3t, t\rangle+|HV\rangle |b_{1}b_{2}\rangle |4t, 2t\rangle \rightarrow 2|HV\rangle |b_{1}b_{2}\rangle D(2t,0)$ \cite{HBSA-wei,BSA3}.
Here $D(2t,0)$ represents that the first photon has a relative time delay of $2t$ compared with the second photon, while $D(0,0)$ means that the two photons have no relative time delay.
Therefore, Eq. (\ref{eq14}) is converted into
\begin{eqnarray}\label{eq17}
\begin{split}
|\Omega_{2}^{\prime}\rangle
=&\frac{\alpha\beta\gamma\delta}{4}
[|\phi_{0}^{++}\rangle_{AB^{\prime}}\otimes(|HH\rangle|b_{1}b_{1}\rangle-|HH\rangle|b_{2}b_{2}\rangle\\&
-|HH\rangle|b_{1}b_{2}\rangle
+|HH\rangle|b_{2}b_{1}\rangle
-|VV\rangle|b_{1}b_{1}\rangle\\&
+|VV\rangle|b_{2}b_{2}\rangle
+|VV\rangle|b_{1}b_{2}\rangle
-|VV\rangle|b_{2}b_{1}\rangle\\&
-|HH\rangle|a_{1^{\prime}}a_{1^{\prime}}\rangle
+|HH\rangle|a_{2^{\prime}}a_{2^{\prime}}\rangle
-|HH\rangle|a_{1^{\prime}}a_{2^{\prime}}\rangle\\&
+|HH\rangle|a_{2^{\prime}}a_{1^{\prime}}\rangle
+|VV\rangle|a_{1^{\prime}}a_{1^{\prime}}\rangle
-|VV\rangle|a_{2^{\prime}}a_{2^{\prime}}\rangle\\&
+|VV\rangle|a_{1^{\prime}}a_{2^{\prime}}\rangle
-|VV\rangle|a_{2^{\prime}}a_{1^{\prime}}\rangle)D(2t,0)\\&
+|\phi_{0}^{--}\rangle_{AB^{\prime}}\otimes(|VH\rangle|b_{1}b_{1}\rangle-|VH\rangle|b_{2}b_{2}\rangle\\&
-|VH\rangle|b_{1}b_{2}\rangle
+|VH\rangle|b_{2}b_{1}\rangle
+|HV\rangle|b_{2}b_{2}\rangle\\&
-|HV\rangle|b_{1}b_{1}\rangle
+|HV\rangle|b_{1}b_{2}\rangle
-|HV\rangle|b_{2}b_{1}\rangle\\&
-|HV\rangle|a_{1^{\prime}}a_{1^{\prime}}\rangle
+|HV\rangle|a_{2^{\prime}}a_{2^{\prime}}\rangle
-|HV\rangle|a_{1^{\prime}}a_{2^{\prime}}\rangle\\&
+|HV\rangle|a_{2^{\prime}}a_{1^{\prime}}\rangle
+|VH\rangle|a_{1^{\prime}}a_{1^{\prime}}\rangle
-|VH\rangle|a_{2^{\prime}}a_{2^{\prime}}\rangle\\&
+|VH\rangle|a_{1^{\prime}}a_{2^{\prime}}\rangle
-|VH\rangle|a_{2^{\prime}}a_{1^{\prime}}\rangle)D(2t,0)\\&
+2|\phi_{0}^{+-}\rangle_{AB^{\prime}}\otimes(|HH\rangle|a_{2^{\prime}}b_{1}\rangle-|HH\rangle|a_{1^{\prime}}b_{2}\rangle\\&
-|VV\rangle|a_{2^{\prime}}b_{1}\rangle+|VV\rangle|a_{1^{\prime}}b_{2}\rangle)D(0,0)\\&
+2|\phi_{0}^{-+}\rangle_{AB^{\prime}}\otimes(|VH\rangle|a_{1^{\prime}}b_{1}\rangle-|HV\rangle|a_{1^{\prime}}b_{1}\rangle\\&
-|VH\rangle|a_{2^{\prime}}b_{2}\rangle+|HV\rangle|a_{2^{\prime}}b_{2}\rangle)D(0,0)],
\end{split}
\end{eqnarray}
where
\begin{eqnarray}\label{eq18}
\begin{split}
&|\phi_{0}^{+\pm}\rangle_{AB^{\prime}}=\frac{1}{2}(|HH\rangle+|VV\rangle) \otimes (|a_{1}b_{1^{\prime}}\rangle\pm|a_{2}b_{2^{\prime}}\rangle),\\
&|\phi_{0}^{\pm+}\rangle_{AB^{\prime}}=\frac{1}{2}(|HH\rangle\pm|VV\rangle) \otimes (|a_{1}b_{1^{\prime}}\rangle+|a_{2}b_{2^{\prime}}\rangle).
\end{split}
\end{eqnarray}
For the recycling state $|\Omega_{3}\rangle$ shown in Eq. (\ref{eq14}), we can infer that after the spatial-based and polarization-based Hadamard operations are applied on photons $A^{\prime}$ and $B$, it is transformed into
\begin{eqnarray}\label{eq19}
\begin{split}
|\Omega_{3}^{\prime}\rangle
=&\frac{\sqrt{(|\alpha|^4+|\beta|^4)(|\gamma|^4+|\delta|^4)}}{2\sqrt{14}}\\&
\times[|\phi_{1}^{++}\rangle_{AB^{\prime}} \otimes
    (|HH\rangle|b_{1}b_{1}\rangle+|HH\rangle|b_{2}b_{2}\rangle\\&
    +|VV\rangle|b_{1}b_{1}\rangle+|VV\rangle|b_{2}b_{2}\rangle\\&-|HH\rangle|a_{1^{\prime}}a_{1^{\prime}}\rangle
    -|HH\rangle|a_{2^{\prime}}a_{2^{\prime}}\rangle\\&-|VV\rangle|a_{1^{\prime}}a_{1^{\prime}}\rangle
    -|VV\rangle|a_{2^{\prime}}a_{2^{\prime}}\rangle)D(0,0)\\&
+2|\phi_{1}^{-+}\rangle_{AB^{\prime}} \otimes
    (|HV\rangle|b_{1}b_{1}\rangle+|HV\rangle|b_{2}b_{2}\rangle\\&
    -|HV\rangle|a_{1^{\prime}}a_{1^{\prime}}\rangle-|HV\rangle|a_{2^{\prime}}a_{2^{\prime}}\rangle)D(0,0)\\&
+2|\phi_{1}^{+-}\rangle_{AB^{\prime}} \otimes
    (|HH\rangle|b_{1}b_{2}\rangle+|VV\rangle|b_{1}b_{2}\rangle\\&
    -|HH\rangle|a_{1^{\prime}}a_{2^{\prime}}\rangle-|VV\rangle|a_{1^{\prime}}a_{2^{\prime}}\rangle)D(0,0)\\&
+2|\phi_{1}^{--}\rangle_{AB^{\prime}} \otimes
    (|HV\rangle|b_{1}b_{2}\rangle+|HV\rangle|b_{2}b_{1}\rangle\\&
    -|HV\rangle|a_{1^{\prime}}a_{2^{\prime}}\rangle-|HV\rangle|a_{2^{\prime}}a_{1^{\prime}}\rangle)D(0,0)],
\end{split}
\end{eqnarray}
where
\begin{eqnarray}\label{eq20}
\begin{split}
|\phi_{1}^{+\pm}\rangle_{AB^{\prime}}=&(\alpha^{\prime}|HV\rangle+\beta^{\prime}|VH\rangle) \otimes
                                           (\gamma^{\prime}|a_{1}b_{2^{\prime}}\rangle \\ &\pm \delta^{\prime}|a_{2}b_{1^{\prime}}\rangle),\\
|\phi_{1}^{\pm+}\rangle_{AB^{\prime}}=&(\alpha^{\prime}|HV\rangle\pm\beta^{\prime}|VH\rangle) \otimes
                                          (\gamma^{\prime}|a_{1}b_{2^{\prime}}\rangle \\ &+ \delta^{\prime}|a_{2}b_{1^{\prime}}\rangle),
\end{split}
\end{eqnarray}
with $\alpha^{\prime}=\frac{\alpha^{2}}{\sqrt{|\alpha|^{4}+|\beta|^{4}}}$, $\beta^{\prime}=\frac{\beta^{2}}{\sqrt{|\alpha|^{4}+|\beta|^{4}}}$,
$\gamma^{\prime}=\frac{\gamma^{2}}{\sqrt{|\gamma|^{4}+|\delta|^{4}}}$,
and $\delta^{\prime}=\frac{\delta^{2}}{\sqrt{|\gamma|^{4}+|\delta|^{4}}}$.

Finally, the polarization and spatial information of photons $A^{\prime}$ and $B$ are measured by the $\text{PBSs}$ and the common single-photon detector $D_i$ $(i=1,2,\cdots,8)$.

For $|\Omega_{0}\rangle$ or $|\Omega_{1}\rangle$, after passing through $\text{BS}_5$, $\text{BS}_{5^{\prime}}$, $\text{HWP}^{22.5^{\circ}}_{1,2}$, and $\text{HWP}^{22.5^{\circ}}_{1^{\prime},2^{\prime}}$, this results in the arbitrary single-photon pair $(D_{i},D_{j})$  $(i,j=1,2,\cdots,8)$ triggered with the time interval of $t$ or $3t$, which means the scheme is failed.


For $|\Omega_{2}^{\prime}\rangle$, as shown by Tab. \ref{Table1}, this results in the arbitrary single-photon pair $(D_{i},D_{j})$ fired with the time interval of $2t$, or Alice and Bob both having one detector triggered without time interval, and $|\Omega_{2}^{\prime}\rangle$ collapses to $|\phi_{0}^{\pm\pm}\rangle_{AB^{\prime}}$.
After some classical feed-forward operations (see Tab. \ref{Table1}) are performed on photon $B^{\prime}$, Alice and Bob obtain the desired maximal hyperentangled Bell state $|\phi_{0}^{++}\rangle_{AB^{\prime}}$ with a success probability of $P_{1}=4|\alpha\beta\gamma\delta|^{2}$.

As for $|\Omega_{3}^{\prime}\rangle$, this results in only one common single-photon detector $D_{i}$ fired or two single-photon detectors from one side, Alice or Bob, fired simultaneously.
Meanwhile, $|\Omega_{2}^{\prime}\rangle$ collapses to one of the states given in Eq. (\ref{eq20}).
After implementing the corresponding feed-forward operations on photon $B^{\prime}$, Alice and Bob obtain the state $|\phi_{1}^{++}\rangle_{AB^{\prime}}$ with a probability of $(|\alpha|^4+|\beta|^4)(|\gamma|^4+|\delta|^4)$.
%


\begin{table*}[htbp]
  \centering
  \caption{The relations between the detection signatures and the classical feed-forward operations to complete the hyper-ECP for hyperentangled Bell state.  The spatial-based feed-forward operation $Z^{S}=|b_{1'}\rangle \langle b_{1'}|-|b_{2'}\rangle \langle b_{2'}| $ can be achieved by using a PS, and the polarization-based
  $Z^{P}=|H\rangle \langle H|-|V\rangle \langle V|$ can be accomplished by employing $\text{HWP}^{0^{\circ}}$. The superscript $2t$ on the detector $D_{i}$ indicates a relative time delay of 2t. $|\phi_{0}^{\pm\pm}\rangle_{AB^{\prime}}$ and $|\phi_{1}^{\pm\pm}\rangle_{AB^{\prime}}$ are the desired and recyclable outcomes, respectively.}\label{Table1}
  \setlength{\tabcolsep}{10mm}
  \begin{tabular}{ccccc}

  \hline   \hline

     &    Single-       & Outcomes                 & Feed-             & Success  \\
Time &       photon     &   of                     & forward           &  proba- \\
interval  &  detectors       &  $A$ and $B^{\prime}$    & on $B^{\prime}$   &       bility \\ \hline
0 & $D_{1}, D_{2}, D_{3}, D_{4}$ & $|\phi_{1}^{++}\rangle_{AB^{\prime}}$  & none & \\ 
  & $D_{5}, D_{6}, D_{7}, D_{8}$ &  &  &\\
  & $(D_{1},D_{2}), (D_{3},D_{4})$ & $|\phi_{1}^{-+}\rangle_{AB^{\prime}}$ & $Z^{P}$ &  $(|\alpha|^4$ \\
  & $(D_{5},D_{6}), (D_{7},D_{8})$ &  &  &$+|\beta|^4)$\\ 
  & $(D_{1},D_{4}), (D_{2},D_{3})$ & $|\phi_{1}^{+-}\rangle_{AB^{\prime}}$ & $Z^{S}$ &  $\times(|\gamma|^4$ \\ 
  & $(D_{5},D_{8}), (D_{6},D_{7})$ &  &  &$+|\delta|^4)$\\ 
  & $(D_{1},D_{3}), (D_{2},D_{4})$ & $|\phi_{1}^{--}\rangle_{AB^{\prime}}$ & $Z^{S},Z^{P}$ &   \\
  & $(D_{5},D_{7}), (D_{6},D_{8})$ &  &  &\\

0 & $(D_{1},D_{8}), (D_{2},D_{7})$ & $|\phi_{0}^{+-}\rangle_{AB^{\prime}}$  & $Z^{S}$ & $4|\alpha\beta\gamma\delta|^{2}$ \\ 
  & $(D_{3},D_{6}), (D_{4},D_{5})$ &  &  &\\
  & $(D_{1},D_{6}), (D_{2},D_{5})$ & $|\phi_{0}^{-+}\rangle_{AB^{\prime}}$ & $Z^{P}$ &\\
  & $(D_{3},D_{8}), (D_{4},D_{7})$ &  &  &\\ 

$2t$ & $(D_{1}^{2t},D_{2}), (D_{1},D_{2}^{2t})$ & $|\phi_{0}^{--}\rangle_{AB^{\prime}}$  & $Z^{S},Z^{P}$ &  \\
     & $(D_{3}^{2t},D_{4}), (D_{3},D_{4}^{2t})$ &  &  &\\
     & $(D_{1}^{2t},D_{3}), (D_{1},D_{3}^{2t})$ &  &  &\\ 
     & $(D_{2}^{2t},D_{4}), (D_{2},D_{4}^{2t})$ &  &  &\\ 
     & $(D_{5}^{2t},D_{6}), (D_{5},D_{6}^{2t})$ &  &  &\\ 
     & $(D_{7}^{2t},D_{8}), (D_{7},D_{8}^{2t})$ &  &  &\\ 
     & $(D_{5}^{2t},D_{7}), (D_{5},D_{7}^{2t})$ &  &  &\\
     & $(D_{6}^{2t},D_{8}), (D_{6},D_{8}^{2t})$ &  &  &\\

$2t$ & $(D_{1}^{2t},D_{1}), (D_{2}^{2t},D_{2})$ & $|\phi_{0}^{++}\rangle_{AB^{\prime}}$  & none &  \\
     & $(D_{3}^{2t},D_{3}), (D_{4}^{2t},D_{4})$ &  &  &\\
     & $(D_{5}^{2t},D_{5}), (D_{6}^{2t},D_{6})$ &  &  &\\ 
     & $(D_{7}^{2t},D_{7}), (D_{8}^{2t},D_{8})$ &  &  &\\ 
     & $(D_{1}^{2t},D_{4}), (D_{1},D_{4}^{2t})$ &  &  &\\ 
     & $(D_{2}^{2t},D_{3}), (D_{2},D_{3}^{2t})$ &  &  &\\ 
     & $(D_{5}^{2t},D_{8}), (D_{5},D_{8}^{2t})$ &  &  &\\
     & $(D_{6}^{2t},D_{7}), (D_{6},D_{7}^{2t})$ &  &  &\\
\hline  \hline

\end{tabular}
\end{table*}


After flipping the polarization and spatial mode of photon $B^{\prime}$ of state $|\phi_{1}^{++}\rangle_{AB^{\prime}}$, it is transformed into
\begin{eqnarray}\label{eq21}
\begin{split}
|\phi_{2}^{++}\rangle_{AB^{\prime}}=&(\alpha^{\prime}|HH\rangle+\beta^{\prime}|VV\rangle) \otimes
                                           (\gamma^{\prime}|a_{1}b_{1^{\prime}}\rangle \\ &+ \delta^{\prime}|a_{2}b_{2^{\prime}}\rangle).
\end{split}
\end{eqnarray}
It is obvious that Eq. (\ref{eq21}) has a similar form to a partially hyperentangled Bell state $(\alpha|HH\rangle+\beta|VV\rangle)\otimes(\gamma|a_{1}b_{1}\rangle+\delta|a_{2}b_{2}\rangle)$ as described in Ref. \cite{hyper-ECP1}.
Then, applying the method in Ref. \cite{hyper-ECP1}, Alice and Bob can further extract the maximal hyperentangled Bell state $|\phi_{0}^{++}\rangle_{AB^{\prime}}$ from $|\phi_{2}^{++}\rangle_{AB^{\prime}}$ with a success probability of $4|\alpha^{\prime}\beta^{\prime}\gamma^{\prime}\delta^{\prime}|^{2}$.
Therefore, this improves the total success probability of our hyper-ECP from $P_{1}$ to
\begin{eqnarray}\label{eq22}
\begin{split}
P_{2}&=P_{1}+ 4|\alpha^{\prime}\beta^{\prime}\gamma^{\prime}\delta^{\prime}|^{2}(|\alpha|^4+|\beta|^4)(|\gamma|^4+|\delta|^4)\\
&=4|\alpha\beta\gamma\delta|^{2}+\frac{4|\alpha\beta\gamma\delta|^{4}}{(|\alpha|^4+|\beta|^4)(|\gamma|^4+|\delta|^4)}.
\end{split}
\end{eqnarray}

\begin{figure} [htbp]
  \centering
  \subfigure[]{
  \includegraphics[width=9cm]{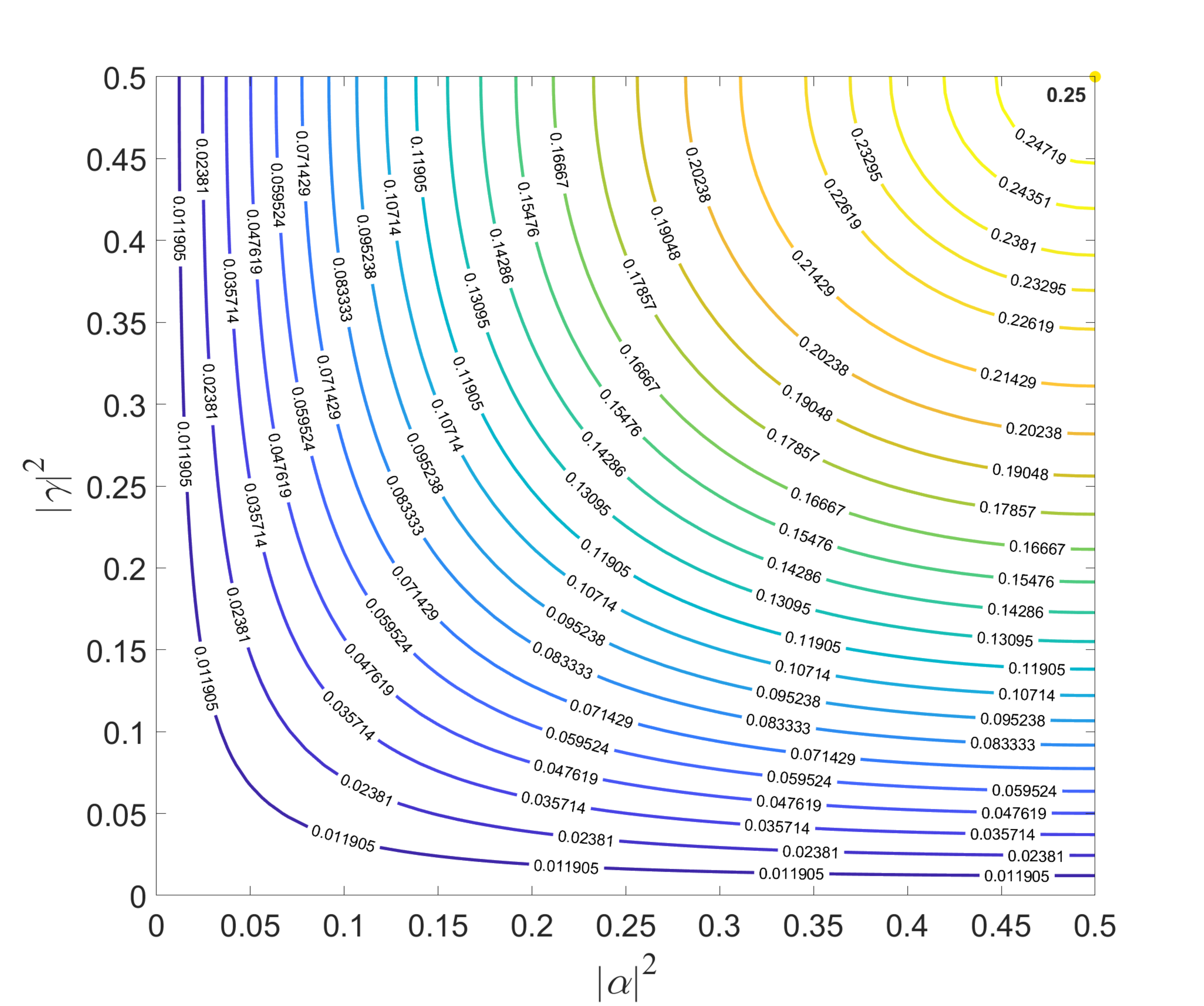}
  \label{a}}
  \subfigure[]{
  \includegraphics[width=9cm]{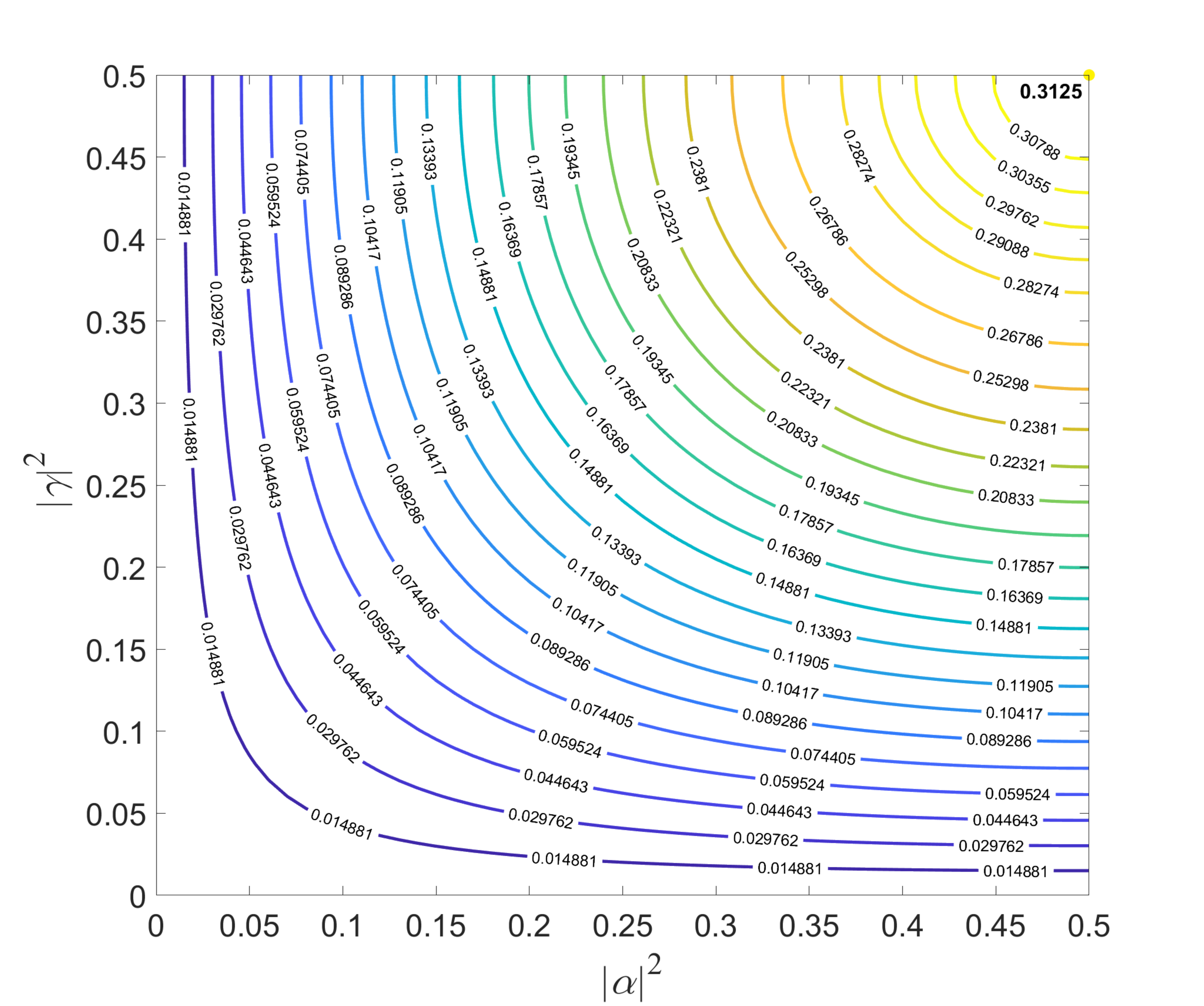}
  \label{b}}
  \caption{The total success probability of the presented hyper-Bell state concentration protocol (a) without and (b) with recycling state $|\phi_{1}^{++}\rangle_{AB^{\prime}}$.}
  \label{Fig.2}
\end{figure}

Figure \ref{Fig.2} plots the success probabilities of $P_{1}$ and $P_{2}$ as a function of $|\alpha|^{2}$ and $|\gamma|^{2}$, where $|\alpha|^{2}=1-|\beta|^{2}\in(0,0.5]$ and $|\gamma|^{2}=1-|\delta|^{2}\in(0,0.5]$ are taken.
As shown in Fig. \ref{a}, $P_{1}=0.25$ without recycling $|\phi_{1}^{++}\rangle_{AB^{\prime}}$, and it is equal to the result shown in Ref. \cite{hyper-ECP1}.
However, after concentrating state $|\phi_{1}^{++}\rangle_{AB^{\prime}}$ again, as shown in Fig. \ref{b}, $P_{1}$ can be increased to $P_{2}=0.3125$ in principle.

\subsection{Hyper-ECP for partially hyperentangled GHZ states with unknown parameters}\label{sec2.2}

\begin{figure} [htbp]
  \centering
  \includegraphics[width=8.6 cm]{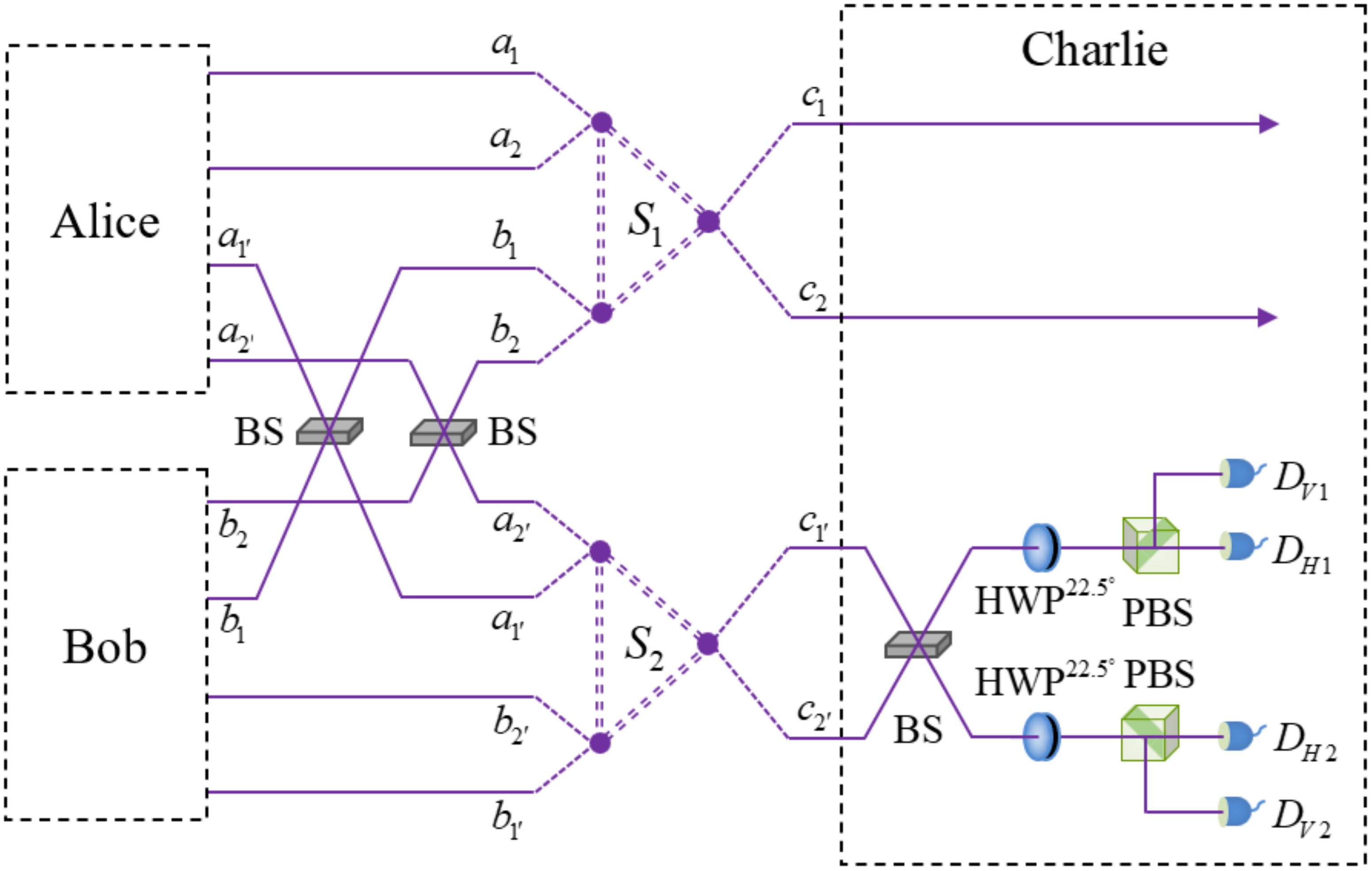}
  \caption{(Color online) Schematic diagram of the hyper-ECP for a hyperentangled GHZ state with unknown parameters.
  $S_{1}$ and $S_{2}$ are partial hyperentanglement sources for $|\psi\rangle_{ABC}$ and $|\psi\rangle_{A^{\prime} B^{\prime} C^{\prime}}$, respectively.}\label{Fig.3}
\end{figure}

Suppose two maximally hyperentangled GHZ states from $S_{1}$ and $S_{2}$ are given by
\begin{eqnarray}\label{eq23}
\begin{split}
|\psi\rangle_{ABC}
=&\frac{1}{2}(|HHH\rangle+|VVV\rangle)
\\&\otimes (|a_{1}b_{1}c_{1}\rangle+|a_{2}b_{2}c_{2}\rangle),
\\
|\psi\rangle_{A^{\prime}B^{\prime}C^{\prime}}
=&\frac{1}{2}(|HHH\rangle+|VVV\rangle)\\ &\otimes (|a_{1^{\prime}}b_{1^{\prime}}c_{1^{\prime}}\rangle+|a_{2^{\prime}}b_{2^{\prime}}c_{2^{\prime}}\rangle).
\end{split}
\end{eqnarray}
%
The $|a_{k}\rangle$ ($|a_{k^{\prime}}\rangle$), $|b_{k}\rangle$ ($|b_{k^{\prime}}\rangle$), and $|c_{k}\rangle$ ($|c_{k^{\prime}}\rangle$), $k=1,2$, are spatial modes of photons $A$ ($A^{\prime}$), $B$ ($B^{\prime}$), and $C$ ($C^{\prime}$), respectively.
As shown in Fig. \ref{Fig.3}, after photon $B$ in the spatial mode $|b_{1}\rangle$ ($|b_{2}\rangle$) and photon $A^{\prime}$ in the spatial mode $|a_{1^\prime}\rangle$ ($|a_{2^\prime}\rangle$) pass through BS, photons $A$, $B$, and $C$ are distributed to three distant parties Alice, Bob, and Charlie, respectively.
Due to the operations of BSs and noisy channels, the six-photon system composed of photons $A$, $B$, $C$, $A'$, $B'$, and $C'$ is given by
\begin{eqnarray}\label{eq24}
\begin{split}
|\Psi_{0}\rangle
=&\frac{1}{\sqrt{2}}(\alpha^{2}|HHHHHH\rangle+\beta^{2}|VVVVVV\rangle)\\
&\otimes  [\gamma^{2}|a_{1}b_{1^{\prime}}c_{1}\rangle(|b_{1}b_{1}\rangle-|a_{1^{\prime}}a_{1^{\prime}}\rangle)|c_{1^{\prime}}\rangle\\
&+\delta^{2}|a_{2}b_{2^{\prime}}c_{2}\rangle(|b_{2}b_{2}\rangle-|a_{2^{\prime}}a_{2^{\prime}}\rangle)|c_{2^{\prime}}\rangle]\\
&+ \frac{\alpha\beta}{2}(|HVHVHV\rangle+|VHVHVH\rangle)\\
&\otimes  [\gamma^{2}|a_{1}b_{1^{\prime}}c_{2}\rangle(|b_{1}b_{1}\rangle-|a_{1^{\prime}}a_{1^{\prime}}\rangle-|b_{1}a_{1^{\prime}}\rangle\\
&+|a_{1^{\prime}}b_{1}\rangle)|c_{1^{\prime}}\rangle+\delta^{2}|a_{2}b_{2^{\prime}}c_{2}\rangle(|b_{2}b_{2}\rangle\\&
-|a_{2^{\prime}}a_{2^{\prime}}\rangle-|b_{2}a_{2^{\prime}}\rangle+|a_{2^{\prime}}b_{2}\rangle)|c_{2^{\prime}}\rangle]\\
&+ \frac{\gamma\delta}{2}(\alpha^{2}|HHHHHH\rangle+\beta^{2}|VVVVVV\rangle\\
&+\alpha\beta|HVHVHV\rangle+\beta\alpha|VHVHVH\rangle)\\
&\otimes  [|a_{1}b_{2^{\prime}}c_{1}\rangle(|b_{1}b_{2}\rangle-|a_{1^{\prime}}a_{2^{\prime}}\rangle-|b_{1}a_{2^{\prime}}\rangle\\
&+|a_{1^{\prime}}b_{2}\rangle)|c_{1^{\prime}}\rangle+|a_{2}b_{1^{\prime}}c_{2}\rangle(|b_{2}b_{1}\rangle\\&
-|a_{2^{\prime}}a_{1^{\prime}}\rangle-|b_{2}a_{1^{\prime}}\rangle+|a_{2^{\prime}}b_{1}\rangle)|c_{2^{\prime}}\rangle].
\end{split}
\end{eqnarray}

The architecture of distilling the maximally hyperentangled GHZ states from $|\Psi_{0}\rangle$ is shown in Fig. \ref{Fig.3}, where Alice and Bob perform the same operations as in Fig. \ref{Fig.1}.
After Alice, Bob, and Charlie have completed concentration operations, similar to that as described in Sec. \ref{sec2}, the relationship between the output states and the corresponding detectors triggered is shown in Tab. \ref{Table2}.
Here, these states in Tab. \ref{Table2} are as follows:
\begin{eqnarray}\label{eq25}
\begin{split}
|\psi^{+\pm}_{0}\rangle_{AB^{\prime}C}
=&\frac{1}{2}(|HHH\rangle+|VVV\rangle)
\\&\otimes (|a_{1}b_{1^{\prime}}c_{1}\rangle\pm|a_{2}b_{2^{\prime}}c_{2}\rangle),\\
|\psi^{\pm+}_{0}\rangle_{AB^{\prime}C}
=&\frac{1}{2}(|HHH\rangle\pm|VVV\rangle)
\\&\otimes (|a_{1}b_{1^{\prime}}c_{1}\rangle + |a_{2}b_{2^{\prime}}c_{2}\rangle).\\
|\psi^{+\pm}_{1}\rangle_{AB^{\prime}C}
=&(\alpha^{\prime}|HVH\rangle+\beta^{\prime}|VHV\rangle)
\\&\otimes (\gamma^{\prime}|a_{1}b_{2^{\prime}}c_{1}\rangle \pm \delta^{\prime}|a_{2}b_{1^{\prime}}c_{2}\rangle),\\
|\psi^{\pm+}_{1}\rangle_{AB^{\prime}C}
=&(\alpha^{\prime}|HVH\rangle\pm\beta^{\prime}|VHV\rangle)
\\&\otimes (\gamma^{\prime}|a_{1}b_{2^{\prime}}c_{1}\rangle+\delta^{\prime}|a_{2}b_{1^{\prime}}c_{2}\rangle).
\end{split}
\end{eqnarray}

\begin{table*}[htbp]
  \centering
  \caption{The relations between the detection signatures, the output states, and the classical feed-forward operations to complete the hyper-ECP for a hyperentangled GHZ state with unknown parameters. $|\psi_{0}^{\pm\pm}\rangle_{AB^{\prime}C}$ and $|\psi_{1}^{\pm\pm}\rangle_{AB^{\prime}C}$ are the desired and recyclable outcomes, respectively.}\label{Table2}
  \begin{tabular}{ccccc}

  \hline   \hline
Detectors      &    Detectors       & Outcomes of                 & Feed-forward               \\
 of Charlie    &  of Alice and Bob     & $A$, $B^{\prime}$, and $C$    & on $B^{\prime}$    \\ \hline
$D_{H1}$ & $D_{1},\,\,\, D_{2},\,\,\, D_{3},\,\,\, D_{4},\,\,\, D_{5},\,\,\, D_{6},\,\,\, D_{7},\,\,\, D_{8}$ & $|\psi_{1}^{++}\rangle_{AB^{\prime}C}$  & none  \\ 
$D_{H2}$ & $(D_{2},D_{3}),\, (D_{1},D_{4}),\,(D_{6},D_{7}),\, (D_{5},D_{8})$ &   \\
$D_{V1}$ & $(D_{1},D_{2}),\, (D_{3},D_{4}),\,(D_{5},D_{6}),\, (D_{7},D_{8})$ &   \\
$D_{V2}$ & $(D_{1},D_{3}),\, (D_{2},D_{4}),\,(D_{5},D_{7}),\, (D_{6},D_{8})$ &    \\

$D_{H2}$ & $D_{1},\,\,\, D_{2},\,\,\, D_{3},\,\,\, D_{4},\,\,\, D_{5},\,\,\, D_{6},\,\,\, D_{7},\,\,\, D_{8}$ & $|\psi_{1}^{+-}\rangle_{AB^{\prime}C}$  & $Z^{S}$  \\
$D_{H1}$ & $(D_{2},D_{3}),\, (D_{1},D_{4}),\,(D_{6},D_{7}),\, (D_{5},D_{8})$ &   \\ 
$D_{V2}$ & $(D_{1},D_{2}),\, (D_{3},D_{4}),\,(D_{5},D_{6}),\, (D_{7},D_{8})$ &   \\
$D_{V1}$ & $(D_{1},D_{3}),\, (D_{2},D_{4}),\,(D_{5},D_{7}),\, (D_{6},D_{8})$ &   \\

$D_{V1}$ & $D_{1},\,\,\, D_{2},\,\,\, D_{3},\,\,\, D_{4},\,\,\, D_{5},\,\,\, D_{6},\,\,\, D_{7},\,\,\, D_{8}$ & $|\psi_{1}^{-+}\rangle_{AB^{\prime}C}$  & $Z^{P}$  \\
$D_{V2}$ & $(D_{2},D_{3}),\, (D_{1},D_{4}),\,(D_{6},D_{7}),\, (D_{5},D_{8})$ &    \\
$D_{H1}$ & $(D_{1},D_{2}),\, (D_{3},D_{4}),\,(D_{5},D_{6}),\, (D_{7},D_{8})$ &   \\ 
$D_{H2}$ & $(D_{1},D_{3}),\, (D_{2},D_{4}),\,(D_{5},D_{7}),\, (D_{6},D_{8})$ &   \\

$D_{V2}$ & $D_{1},\,\,\, D_{2},\,\,\, D_{3},\,\,\, D_{4},\,\,\, D_{5},\,\,\, D_{6},\,\,\, D_{7},\,\,\, D_{8}$ &  $|\psi_{1}^{--}\rangle_{AB^{\prime}C}$  & $Z^{S},Z^{P}$  \\
$D_{V1}$ & $(D_{2},D_{3}),\, (D_{1},D_{4}),\,(D_{6},D_{7}),\, (D_{5},D_{8})$ &   \\
$D_{H2}$ & $(D_{1},D_{2}),\, (D_{3},D_{4}),\,(D_{5},D_{6}),\, (D_{7},D_{8})$ &   \\
$D_{H1}$ & $(D_{1},D_{3}),\, (D_{2},D_{4}),\,(D_{5},D_{7}),\, (D_{6},D_{8})$ &   \\

$D_{H2}$ & $(D_{1},D_{8}),\, (D_{2},D_{7}),\,(D_{3},D_{6}),\, (D_{4},D_{5})$ & $|\psi_{0}^{++}\rangle_{AB^{\prime}C}$ & none   \\
$D_{V1}$ & $(D_{1},D_{6}),\, (D_{2},D_{5}),\,(D_{3},D_{8}),\, (D_{4},D_{7})$ &  &    \\
$D_{H1}$
                      & $(D_{1}^{2t},D_{1}),\,(D_{2}^{2t},D_{2}),\,(D_{3}^{2t},D_{3}),\,(D_{4}^{2t},D_{4}),\,(D_{1}^{2t},D_{4}),\,(D_{2}^{2t},D_{3}),\,
                      (D_{1},D_{4}^{2t}),\,(D_{2},D_{3}^{2t})$ &  &    \\
                      & $(D_{5}^{2t},D_{5}),\,(D_{6}^{2t},D_{6}),\,(D_{7}^{2t},D_{7}),\,(D_{8}^{2t},D_{8}),\,(D_{5}^{2t},D_{8}),\,(D_{6}^{2t},D_{7}),\,
                      (D_{5},D_{8}^{2t}),\,(D_{6},D_{7}^{2t})$ &  &    \\
$D_{V2}$
                      & $(D_{1}^{2t},D_{2}),\,(D_{3}^{2t},D_{4}),\,(D_{1}^{2t},D_{3}),\,(D_{2}^{2t},D_{4}),\,(D_{1},D_{2}^{2t}),\,(D_{3},D_{4}^{2t}),\,
                      (D_{1},D_{3}^{2t}),\,(D_{2},D_{4}^{2t})$ &  &  \\
                      & $(D_{5}^{2t},D_{6}),\,(D_{7}^{2t},D_{8}),\,(D_{5}^{2t},D_{7}),\,(D_{6}^{2t},D_{8}),\,(D_{5},D_{6}^{2t}),\,(D_{7},D_{8}^{2t}),\,
                      (D_{5},D_{7}^{2t}),\,(D_{6},D_{8}^{2t})$ &  &  \\

$D_{H1}$ & $(D_{1},D_{8}),\, (D_{2},D_{7}),\,(D_{3},D_{6}),\, (D_{4},D_{5})$ & $|\psi_{0}^{+-}\rangle_{AB^{\prime}C}$  & $Z^{S}$  \\ 
$D_{V2}$ & $(D_{1},D_{6}),\, (D_{2},D_{5}),\,(D_{3},D_{8}),\, (D_{4},D_{7})$ &  &  \\
$D_{H2}$
                      & $(D_{1}^{2t},D_{1}),\,(D_{2}^{2t},D_{2}),\,(D_{3}^{2t},D_{3}),\,(D_{4}^{2t},D_{4}),\,(D_{1}^{2t},D_{4}),\,(D_{2}^{2t},D_{3}),\,
                      (D_{1},D_{4}^{2t}),\,(D_{2},D_{3}^{2t})$ &  &    \\
                      & $(D_{5}^{2t},D_{5}),\,(D_{6}^{2t},D_{6}),\,(D_{7}^{2t},D_{7}),\,(D_{8}^{2t},D_{8}),\,(D_{5}^{2t},D_{8}),\,(D_{6}^{2t},D_{7}),\,
                      (D_{5},D_{8}^{2t}),\,(D_{6},D_{7}^{2t})$ &  &    \\
$D_{V1}$
                      & $(D_{1}^{2t},D_{2}),\,(D_{3}^{2t},D_{4}),\,(D_{1}^{2t},D_{3}),\,(D_{2}^{2t},D_{4}),\,(D_{1},D_{2}^{2t}),\,(D_{3},D_{4}^{2t}),\,
                      (D_{1},D_{3}^{2t}),\,(D_{2},D_{4}^{2t})$ &  &  \\
                      & $(D_{5}^{2t},D_{6}),\,(D_{7}^{2t},D_{8}),\,(D_{5}^{2t},D_{7}),\,(D_{6}^{2t},D_{8}),\,(D_{5},D_{6}^{2t}),\,(D_{7},D_{8}^{2t}),\,
                      (D_{5},D_{7}^{2t}),\,(D_{6},D_{8}^{2t})$ &  &  \\

$D_{V2}$ & $(D_{1},D_{8}),\, (D_{2},D_{7}),\,(D_{3},D_{6}),\, (D_{4},D_{5})$ &  $|\psi_{0}^{-+}\rangle_{AB^{\prime}C}$  & $Z^{P}$  \\
$D_{H1}$ & $(D_{1},D_{6}),\, (D_{2},D_{5}),\,(D_{3},D_{8}),\, (D_{4},D_{7})$ &  & \\ 
$D_{V1}$
                      & $(D_{1}^{2t},D_{1}),\,(D_{2}^{2t},D_{2}),\,(D_{3}^{2t},D_{3}),\,(D_{4}^{2t},D_{4}),\,(D_{1}^{2t},D_{4}),\,(D_{2}^{2t},D_{3}),\,
                      (D_{1},D_{4}^{2t}),\,(D_{2},D_{3}^{2t})$ &  &    \\
                      & $(D_{5}^{2t},D_{5}),\,(D_{6}^{2t},D_{6}),\,(D_{7}^{2t},D_{7}),\,(D_{8}^{2t},D_{8}),\,(D_{5}^{2t},D_{8}),\,(D_{6}^{2t},D_{7}),\,
                      (D_{5},D_{8}^{2t}),\,(D_{6},D_{7}^{2t})$ &  &    \\
$D_{H2}$
                      & $(D_{1}^{2t},D_{2}),\,(D_{3}^{2t},D_{4}),\,(D_{1}^{2t},D_{3}),\,(D_{2}^{2t},D_{4}),\,(D_{1},D_{2}^{2t}),\,(D_{3},D_{4}^{2t}),\,
                      (D_{1},D_{3}^{2t}),\,(D_{2},D_{4}^{2t})$ &  &  \\
                      & $(D_{5}^{2t},D_{6}),\,(D_{7}^{2t},D_{8}),\,(D_{5}^{2t},D_{7}),\,(D_{6}^{2t},D_{8}),\,(D_{5},D_{6}^{2t}),\,(D_{7},D_{8}^{2t}),\,
                      (D_{5},D_{7}^{2t}),\,(D_{6},D_{8}^{2t})$ &  &  \\

$D_{V1}$ & $(D_{1},D_{8}),\, (D_{2},D_{7}),\,(D_{3},D_{6}),\, (D_{4},D_{5})$& $|\psi_{0}^{--}\rangle_{AB^{\prime}C}$ & $Z^{S},Z^{P}$  \\
$D_{H2}$ & $(D_{1},D_{6}),\, (D_{2},D_{5}),\,(D_{3},D_{8}),\, (D_{4},D_{7})$& & \\
$D_{V2}$
                      & $(D_{1}^{2t},D_{1}),\,(D_{2}^{2t},D_{2}),\,(D_{3}^{2t},D_{3}),\,(D_{4}^{2t},D_{4}),\,(D_{1}^{2t},D_{4}),\,(D_{2}^{2t},D_{3}),\,
                      (D_{1},D_{4}^{2t}),\,(D_{2},D_{3}^{2t})$ &  &    \\
                      & $(D_{5}^{2t},D_{5}),\,(D_{6}^{2t},D_{6}),\,(D_{7}^{2t},D_{7}),\,(D_{8}^{2t},D_{8}),\,(D_{5}^{2t},D_{8}),\,(D_{6}^{2t},D_{7}),\,
                      (D_{5},D_{8}^{2t}),\,(D_{6},D_{7}^{2t})$ &  &    \\
$D_{H1}$
                      & $(D_{1}^{2t},D_{2}),\,(D_{3}^{2t},D_{4}),\,(D_{1}^{2t},D_{3}),\,(D_{2}^{2t},D_{4}),\,(D_{1},D_{2}^{2t}),\,(D_{3},D_{4}^{2t}),\,
                      (D_{1},D_{3}^{2t}),\,(D_{2},D_{4}^{2t})$ &  &  \\
                      & $(D_{5}^{2t},D_{6}),\,(D_{7}^{2t},D_{8}),\,(D_{5}^{2t},D_{7}),\,(D_{6}^{2t},D_{8}),\,(D_{5},D_{6}^{2t}),\,(D_{7},D_{8}^{2t}),\,
                      (D_{5},D_{7}^{2t}),\,(D_{6},D_{8}^{2t})$ &  &  \\

\hline  \hline

\end{tabular}
\end{table*}

After performing the feed-forward operations on photon $B^{\prime}$ shown in Tab. \ref{Table2}, Alice, Bob, and Charlie can obtain the maximally hyperentangled GHZ state $|\psi^{++}_{0}\rangle_{AB^{\prime}C}$ with a success probability of $4|\alpha\beta\gamma\delta|^{2}$ or a recyclable hyperentangled state $|\psi^{++}_{1}\rangle_{AB^{\prime}C}$ with a probability of $(|\alpha|^4+|\beta|^4)(|\gamma|^4+|\delta|^4)$.
It is easy to verify that the total success probability of the scheme shown in Fig. \ref{Fig.3} is equal to that of the one shown in Fig. \ref{Fig.2} by recycling $|\psi^{++}_{1}\rangle_{AB^{\prime}C}$.
Additionally, as long as the arbitrary single-photon pair $(D_{i},D_{j})$ $(i,j=1,2,\cdots,8)$ belonging to Alice and Bob is triggered with a time interval of $t$ or $3t$, the concentration process fails and is terminated, regardless of the detection signatures of Charlie.

\section{Discussion and Summary}\label{sec3}

Hyperentanglement concentration is one of the methods to decrease the influence of channel noise in high-capacity long-distance quantum communication.
ECP or hyper-ECP is used to deal with the case where the pure state remains after being subjected to noisy channels.
In practice, when affected by noise, the input state is not necessarily still pure but may be a mixed state, in which case EPPs are required.
In this article, we design a hyper-ECP for unknown hyperentangled Bell states in both the polarization and spatial DOFs that can be unambiguously heralded by the detection signatures. Remarkably, this heralded hyper-ECP can be extended to concentrate the GHZ state case.

In the constructions of previous hyper-ECPs using linear optics, PBSs are employed to complete the parity-check measurement of polarization photon pairs \cite{hyper-ECP1,hyper-ECP2,hyper-ECP3,hyper-ECP4}.
Meanwhile, postselection is necessary to explicitly identify the instances where each of the spatial modes contains exactly one photon without destruction of the incident photon, which leads to schemes that cannot be accomplished simply with linear optical elements.
We avoid postselections or sophisticated photon-number-resolving detectors by introducing time-delay DOF as in Ref. \cite{BSA3}, and by the detection signatures of the common single-photon detector pair $(D_{i},D_{j})$, the instances
$\{|HH\rangle|a_{1}b_{1^{\prime}}\rangle$, $|HH\rangle|a_{2}b_{2^{\prime}}\rangle$, $|VV\rangle|a_{1}b_{1^{\prime}}\rangle$, $|VV\rangle|a_{2}b_{2^{\prime}}\rangle\}$, $\{|HV\rangle|a_{1}b_{2^{\prime}}\rangle$, $|HV\rangle|a_{2}b_{1^{\prime}}\rangle$, $|VH\rangle|a_{1}b_{2^{\prime}}\rangle$, $|VH\rangle|a_{2}b_{1^{\prime}}\rangle\}$, and the other terms can be perfectly distinguished without destruction of the incident photon.
Additionally, the success probability of our linear optic architectures can be increased from $\frac{1}{4}$  to $\frac{5}{16}$ by recycling partially hyperentangled states.
In previous works, only hyper-ECPs assisted by nonlinear mediates allow the recyclable program to efficiently improve the success probability \cite{hyper-ECP5,hyper-ECP6,hyper-ECP7,hyper-ECP8}.
That is attributed to the quantum anti-Zeno effect, which accelerates the dynamic evolution towards the target state by frequent measurement \cite{Zeno1}.
Although the success probability of $\frac{5}{16}$ is still lower than that of nonlinear hyper-ECPs, linear optic implementations of our schemes are readily available in practice with current technology.
In the overall process, we assume that the performances of linear optical elements and single-photon detectors are perfect.
However, this does not work in a practical experiment, and the schemes may be negatively affected by noise during application \cite{Wang2018}.
Quantum Zeno effect can enhance the measurement accuracy of entangled probes to resist these disadvantages \cite{Zeno2}.

In summary, we present a practically enhanced hyper-ECP for the hyperentangled Bell state in both the polarization and spatial mode DOFs with unknown parameters using available linear optical elements and common single-photon detectors.
By introducing the time-delay DOF to unbalanced interferences, the practicality of our protocol is efficiently enhanced as the scheme can be perfectly heralded by the unique detection signatures, and postselection or sophisticated photon-number-resolving detectors are not needed.
Moreover, the success probability of our linear optical hyper-ECP is higher than that of Ref. \cite{hyper-ECP1} by recycling partially hyperentangled states.
%
%
Note that different from previous works, the hyperparallel parity-check measurements in our scheme acting on polarization and spatial
DOFs are simultaneously accomplished in one step.
We also extend the hyper-ECP for hyperentangled Bell states to generic  hyperentangled GHZ states.
We will study nonpostselection EPP in the future.

\medskip

\section*{ACKNOWLEDGEMENTS} \par

This work is supported by the Fundamental Research Funds for the Central Universities under Grant No. FRF-TP-19-011A3 and the National Natural Science Foundation of China under Grant No. 11604012.


\medskip



\begin{thebibliography}{99}
\bibitem{1}M. A. Nielsen and I. L. Chuang, \emph{Quantum Computation and Quantum Information} (Cambridge University Press, Cambridge, 2000). 

\bibitem{2}D. P. DiVincenzo, Quantum computation, Science 270, 255 (1995). 

\bibitem{3}A. K. Ekert, Quantum cryptography based on Bell's theorem, Phys. Rev. Lett. 67, 661 (1991).

\bibitem{4}X. F. Ma, P. Zeng, and H. Y. Zhou, Phase-Matching Quantum Key Distribution, Phys. Rev. X 8, 031043 (2018).

\bibitem{5}S. T. Ren, Y. Wang, and X. L. Su, Hybrid quantum key distribution network, Sci. China-Inf. Sci. 65, 200502 (2022).

\bibitem{6}X. L. Wang, X. D. Cai, Z. E. Su, M. C. Chen, D. Wu, L. Li, N. L. Liu, C. Y. Lu, and J. W. Pan, Quantum teleportation of multiple degrees of freedom of a single photon, Nature (London) 518, 516 (2015).



\bibitem{7}P. Lipka-Bartosik and P. Skrzypczyk, Catalytic Quantum Teleportation, Phys. Rev. Lett. 127, 080502 (2021).


\bibitem{8}C. H. Bennett and S. J. Wiesner, Communication via one- and two-particle operators on Einstein-Podolsky-Rosen states, Phys. Rev. Lett. 69, 2881 (1992).



\bibitem{9}Y. X. Chen, S. S Liu, Y. B. Lou, and J. T. Jing, Orbital Angular Momentum Multiplexed Quantum Dense Coding, Phys. Rev. Lett. 127, 093601 (2021).

\bibitem{10}J. Bogdanski, N. Rafiei, and M. Bourennane, Experimental quantum secret sharing using telecommunication fiber, Phys. Rev. A 78, 062307 (2008).

\bibitem{11}K. Senthoor and P. K. Sarvepalli, Communication efficient quantum secret sharing, Phys. Rev. A 100, 052313 (2019).

\bibitem{12}S. M. Lee, S. W. Lee, H. Jeong, and H. S. Park, Quantum Teleportation of Shared Quantum Secret, Phys. Rev. Lett. 124, 060501 (2020).

\bibitem{13}W. Zhang, D. S. Ding, Y. B. Sheng, L. Zhou, B. S. Shi, and G. C. Guo, Quantum Secure Direct Communication with Quantum Memory, Phys. Rev. Lett. 118, 220501 (2017).

\bibitem{14}Z. W. Cao, L. Wang, K. X. Liang, G. Chai, and J. Y. Peng, Continuous-Variable Quantum Secure Direct Communication Based on Gaussian Mapping, Phys. Rev. Applied 16, 024012 (2021).

\bibitem{15}Y. B. Sheng, L. Zhou, and G. L. Long, One-step quantum secure direct communication, Sci. Bull. 67, 367 (2022).

\bibitem{single-qubit} B. N. Simon, C. M. Chandrashekar, and S. Simon, Hamilton's turns as a visual tool kit for designing single-qubit unitary gates, Phys. Rev. A 85, 022323 (2012).

\bibitem{17}T. J. Wang, Y. Lu, and G. L. Long, Generation and complete analysis of the hyperentangled Bell state for photons assisted by quantum-dot spins in optical microcavities, Phys. Rev. A 86, 042337 (2012).

\bibitem{18}J. T. Barreiro, N. K. Langford, N. A. Peters, and P. G. Kwiat, Generation of Hyperentangled Photon Pairs, Phys. Rev. Lett. 95, 260501 (2005).

\bibitem{momentum1}M. Barbieri, F. D. Martini, P. Mataloni, G. Vallone, and A. Cabello, Enhancing the Violation of the Einstein-Podolsky-Rosen Local Realism by Quantum Hyperentanglement, Phys. Rev. Lett. 97, 140407 (2006).

\bibitem{momentum2}M. Barbieri, G. Vallone, P. Mataloni, and F. D. Martini, Complete and deterministic discrimination of polarization Bell states assisted by momentum entanglement, Phys. Rev. A 75, 042317 (2007).

\bibitem{OAM1}A. Mair, A. Vaziri, G. Weihs, and A. Zeilinger, Entanglement of the orbital angular momentum states of photons, Nature (London) 412, 313 (2001).

\bibitem{OAM2}D. Bhatti, J. von Zanthier, and G. S. Agarwal, Entanglement of polarization and orbital angular momentum, Phys. Rev. A 91, 062303 (2015).


\bibitem{OAM3}J. T. Barreiro, T. C. Wei, and P. G. Kwiat, Beating the channel capacity limit for linear photonic superdense coding, Nat. Phys. 4, 282 (2008).

\bibitem{HPQC1}B. C. Ren, G. Y. Wang, and F. G. Deng, Universal hyperparallel hybrid photonic quantum gates with dipoleinduced transparency in the weak-coupling regime, Phys. Rev. A 91, 032328 (2015).

\bibitem{HPQC2}T. Li and G. L. Long, Hyperparallel optical quantum computation assisted by atomic ensembles embedded in double-sided optical cavities, Phys. Rev. A 94, 022343 (2016).

\bibitem{HPQC3}B. C. Ren and F. G. Deng, Robust hyperparallel photonic quantum entangling gate with cavity QED, Opt. Express 25, 10863 (2017).

\bibitem{HBSA1}T. C. Wei, J. T. Barreiro, and P. G. Kwiat, Hyperentangled Bell-state analysis, Phys. Rev. A 75, 060305 (2007).

\bibitem{HBSA2}Y. B. Sheng, F. G. Deng, and G. L. Long, Complete hyperentangled-Bell-state analysis for quantum communication, Phys. Rev. A 82, 032318 (2010).


\bibitem{HBSA3}Q. Liu and M. Zhang, Generation and complete nondestructive analysis of hyperentanglement assisted by nitrogen-vacancy centers in resonators, Phys. Rev. A 91, 062321 (2015).

\bibitem{HBSA4}G. Y. Wang, Q. Ai, B. C. Ren, T. Li, and F. G. Deng, Error-detected generation and complete analysis of hyperentangled Bell states for photons assisted by quantum-dot spins in double-sided optical microcavities, Opt. Express 24, 28444 (2016).

\bibitem{HBSA-wei}X. J. Zhou, W. Q. Liu, H. R. Wei, Y. B. Zheng, and Fang-Fang Du, Deterministic and complete hyperentangled Bell states analysis assisted by frequency and time interval degrees of freedom, Front. Phys. (Beijing) 17, 41502 (2022).

\bibitem{HEPP1}G. Y. Wang, Q. Liu, and F. G. Deng, Hyperentanglement purification for two-photon six-qubit quantum systems, Phys. Rev. A 94, 032319 (2016).

\bibitem{HEPP2}T. J. Wang, S. C. Mi, and C. Wang, Hyperentanglement purification using imperfect spatial entanglement, Opt. Express 25, 2969 (2017).

\bibitem{HEPP3}F. F. Du, Y. T. Liu, Z. R. Shi, Y. X. Liang, J. Tang, and J. Liu, Efficient hyperentanglement purification for three-photon systems with the fidelity-robust quantum gates and hyperentanglement link, Opt. Express 27, 27046 (2019).

\bibitem{BSA1}S. P. Walborn, S. P\'{a}dua, and C. H. Monken, Hyperentanglement-assisted Bell-state analysis, Phys. Rev. A 68, 042313 (2003).

\bibitem{BSA2}C. Schuck, G. Huber, C. Kurtsiefer, and H. Weinfurter, Complete Deterministic Linear Optics Bell State Analysis, Phys. Rev. Lett. 96, 190501 (2006).

\bibitem{BSA3}B. P. Williams, R. J. Sadlier, and T. S. Humble, Superdense Coding over Optical Fiber Links with Complete Bell-State Measurements, Phys. Rev. Lett. 118, 050501 (2017).


\bibitem{EPP1}C. H. Bennett, G. Brassard, S. Popescu, B. Schumacher, J. A. Smolin, and W. K. Wootters, Purification of Noisy Entanglement and Faithful Teleportation via Noisy Channels, Phys. Rev. Lett. 76, 722 (1996).

\bibitem{EPP2}Y. B. Sheng and F. G. Deng, One-step deterministic polarization-entanglement purification using spatial entanglement, Phys. Rev. A 82, 044305 (2010).

\bibitem{EPP3}X. H. Li, Deterministic polarization-entanglement purification using spatial entanglement, Phys. Rev. A 82, 044304 (2010).

\bibitem{EPP4}F. Riera-S\`abat, P. Sekatski, A. Pirker, and W. D\"ur, Entanglement-Assisted Entanglement Purification, Phys. Rev. Lett. 127, 040502 (2021).

\bibitem{EPP5}C. X. Huang, X. M. Hu, B. H. Liu, L. Zhou, Y. B. Sheng, C. F. Li, and G. C. Guo, Experimental one-step deterministic polarization entanglement purification, Sci. Bull. 67, 593 (2022).



\bibitem{ECP1}C. H. Bennett, H. J. Bernstein, S. Popescu, and B. Schumacher, Concentrating partial entanglement by local operations, Phys. Rev. A 53, 2046 (1996).


\bibitem{ECP2}T. Yamamoto, M. Koashi, and N. Imoto, Concentration and purification scheme for two partially entangled photon pairs, Phys. Rev. A 64, 012304 (2001).

\bibitem{ECP2-experiment1}Z. Zhao, T. Yang, Y. A. Chen, A. N. Zhang, and J. W. Pan, Experimental Realization of Entanglement Concentration and a Quantum Repeater, Phys. Rev. Lett. 90, 207901 (2003).


\bibitem{ECP3}M. Yang, Y. Zhao, W. Song, and Z. L. Cao, Entanglement concentration for unknown atomic entangled states via entanglement swapping, Phys. Rev. A 71, 044302 (2005).

\bibitem{ECP4}Y. B. Sheng, F. G. Deng, and H. Y. Zhou, Nonlocal entanglement concentration scheme for partially entangled multipartite systems with nonlinear optics, Phys. Rev. A 77, 062325 (2008).


\bibitem{ECP5}Y. B. Sheng, L. Zhou, and S. M. Zhao, Efficient two-step entanglement concentration for arbitrary $W$ states, Phys. Rev. A 85, 042302 (2012).


\bibitem{ECP6}H. Zhang and H. B. Wang, Entanglement concentration of microwave photons based on the Kerr effect in circuit QED, Phys. Rev. A 95, 052314 (2017).

\bibitem{ECP7}S. S. Chen, H. Zhang, Q. Ai, and G. J. Yang, Phononic entanglement concentration via optomechanical interactions, Phys. Rev. A 100, 052306 (2019).






\bibitem{hyper-ECP1}B. C. Ren, F. F. Du, and F. G. Deng, Hyperentanglement concentration for two-photon four-qubit systems with linear optics, Phys. Rev. A 88, 012302 (2013).

\bibitem{hyper-ECP2}X. H. Li and S. Ghose, Hyperentanglement concentration for time-bin and polarization hyperentangled photons, Phys. Rev. A 91, 062302 (2015).

\bibitem{hyper-ECP3}B. C. Ren, H. Wang, F. Alzahrani, A. Hobiny, and F. G. Deng, Hyperentanglement concentration of nonlocal two photon six-qubit systems with linear optics, Ann. Phys. (New York) 385, 86 (2017).

\bibitem{hyper-ECP4}H. Wang, B. C. Ren, A. H. Wang, A. Alsaedi, T. Hayat, and F. G. Deng, General hyperentanglement concentration for polarization-spatial-time-bin multi-photon systems with linear optics, Front. Phys. (Beijing) 13, 130315 (2018).


\bibitem{hyper-ECP02}C. Y. Li and Y. Shen, Asymmetrical hyperentanglement concentration for entanglement of polarization and orbital angular momentum, Opt. Express 27, 13172 (2019).


\bibitem{hyper-ECP5}X. H. Li and S. Ghose, Efficient hyperconcentration of nonlocal multipartite entanglement via the cross-Kerr nonlinearity, Opt. Express 23, 3550 (2015).

\bibitem{hyper-ECP6}B. C. Ren and G. L. Long, General hyperentanglement concentration for photon systems assisted by quantum dot spins inside optical microcavities, Opt. Express 22, 6547 (2014).

\bibitem{hyper-ECP7}B. C. Ren and F. G. Deng, Hyperentanglement purication and concentration assisted by diamond NV centers inside photonic crystal cavities, Laser Phys. Lett. 10, 115201 (2013).

\bibitem{hyper-ECP8}B. C. Ren and G. L. Long, Highly efficient hyperentanglement concentration with two steps assisted by quantum swap gates, Sci. Rep. 5, 16444 (2015).




\bibitem{Zeno1}Q. Ai, Y. Li, H. Zheng, and C. P. Sun, Quantum anti-Zeno effect without rotating wave approximation, Phys. Rev. A 81, 042116 (2010).

\bibitem{Wang2018}B. X. Wang, M. J. Tao, Q. Ai, T. Xin, N. Lambert, D. Ruan, Y. C. Cheng, F. Nori, F. G. Deng, and G. L. Long, Efficient quantum simulation of photosynthetic light harvesting, npj Quantum Inf. 4, 52 (2018).

\bibitem{Zeno2}X. Y. Long, W. T. He, N. N. Zhang, K. Tang, Z. D. Lin, H. F. Liu, X. F. Nie, G. R. Feng, J. Li, T. Xin, Q. Ai, and D. W. Lu, Entanglement-Enhanced Quantum Metrology in Colored Noise by Quantum Zeno Effect, Phys. Rev. Lett. 129, 070502 (2022).

\end{thebibliography}
\end{document}